\documentclass[aps,prl,showpacs,twocolumn,superscriptaddress]{revtex4}
\usepackage{amsmath}
\usepackage{amssymb}
\usepackage{epsfig}
\usepackage{graphicx,amsmath}
\begin{document}

\title{Energy scales and magnetoresistance at a quantum critical point}

\author{V.R. Shaginyan}\email{vrshag@thd.pnpi.spb.ru}
\affiliation{Petersburg Nuclear Physics Institute, RAS, Gatchina,
188300, Russia}\affiliation{Racah Institute of Physics, Hebrew
University, Jerusalem 91904, Israel} \affiliation{CTSPS, Clark
Atlanta University, Atlanta, Georgia 30314, USA}
\author{M.Ya. Amusia}\affiliation{Racah Institute
of Physics, Hebrew University, Jerusalem 91904, Israel}
\author{A.Z. Msezane}
\affiliation{CTSPS, Clark Atlanta University, Atlanta, Georgia
30314, USA}
\author{K.G. Popov}
\affiliation{Komi Science Center, Ural Division, RAS, 3a, Chernova
street Syktyvkar, 167982, Russia}
\author{V.A. Stephanovich}
\affiliation{Opole University, Institute of Mathematics and
Informatics, Opole, 45-052, Poland}

\begin{abstract}

The magnetoresistance (MR) of $\rm CeCoIn_5$ is notably different
from that in many conventional metals. We show that a pronounced
crossover from negative to positive MR at elevated temperatures and
fixed magnetic fields is determined by the scaling behavior of
quasiparticle effective mass. At a quantum critical point (QCP)
this dependence generates kinks (crossover points from fast to slow
growth) in thermodynamic characteristics (like specific heat,
magnetization etc) at some temperatures when a strongly correlated
electron system transits from the magnetic field induced Landau
Fermi liquid (LFL) regime to the non-Fermi liquid (NFL) one taking
place at rising temperatures. We show that the above kink-like
peculiarity separates two distinct energy scales in QCP vicinity -
low temperature LFL scale and high temperature one related to NFL
regime. Our comprehensive theoretical analysis of experimental data
permits to reveal for the first time new MR and kinks scaling
behavior as well as to identify the physical reasons for above
energy scales.

\end{abstract}


\maketitle

\section{Introduction}

An explanation of rich and striking behavior of strongly correlated
electron system in heavy fermion (HF) metals is, as years before,
among the main problems  of condensed matter physics. One of the
most interesting and puzzling issues in the research of HF metals
is their anomalous normal-state transport properties. Measurements
of magnetoresistance (MR) on $\rm{ CeCoIn_5}$\cite{pag,mal} have
shown that it is notably different from ordinary weak-field orbital
MR described by Kohler's rule which holds in many conventional
metals, see e.g. \cite{zim}. At fixed magnetic fields $B$, MR of
$\rm CeCoIn_5$ exhibits a crossover from negative (low
temperatures) to positive (hight temperatures) one at temperature
growth \cite{pag,mal}. This crossover is hard to explain within
both conventional Fermi liquid approach for metals and in terms of
Kondo systems \cite{daybell}. To explain this effect, it has been
assumed that the crossover can be attributed to some distinct
energy scales revealed by kinks (crossover points from fast to slow
growth) in thermodynamic characteristics (like specific heat,
magnetization etc) and leading to a change of spin fluctuations
character with increasing of the applied magnetic field strength
\cite{pag,mal,daybell,loh,si,steg,kon,nak}.

Here we investigate the NFL-LFL transition region (we call it below
crossover region), where MR changes its sign. The modified Kohler's
rule (MR versus tangent of Hall angle) have been utilized to
describe MR data \cite{kon,nak}. In this region, both Kohler's rule
and its modified version do not work. In Landau Fermi liquid (LFL)
regime, the quasiparticles were observed in measurements of
transport properties in $\rm CeCoIn_5$ \cite{pag1}. An analysis of
above thermodynamic quantities shows that quasiparticles exist in
both LFL and crossover regimes when strongly correlated Fermi
systems like HF metals or two-dimensional (2D) $\rm ^3He$
\cite{ckz,obz,shag1,shag2,shag3,khodb} transit from LFL to NFL
behavior. It is of crucial importance to verify whether
quasiparticles with effective mass $M^*$ still exist and determine
the transport properties and energy scales in HF metals in
crossover region. On the other hand, even early measurements on HF
metals gave evidences in favor of the quasiparticles existence. For
example, the application of magnetic field $B$ restores LFL
behavior of HF metals which demonstrate NFL properties in the
absence of the field . In that case the empirical Kadowaki-Woods
(KW) ratio $K$ is conserved, $K=A(B)/\gamma_0^2(B)\propto
A(B)/\chi^2(B)=const$ \cite{kadw,kh_z,tky} where $\gamma_0=C/T$,
$C$ is a heat capacity, $\chi$ is a magnetic susceptibility and
$A(B)$ is a coefficient determining the temperature dependence of
the resistivity $\rho=\rho_0+A(B)T^2$. Here $\rho_0$ is the
residual resistance. The observed conservation of $K$ can be hardly
interpreted within scenarios when quasiparticles are suppressed,
for there is no reason to expect that $\gamma_0(B)$, $\chi(T)$,
$A(B)$ and other transport and thermodynamic quantities like
thermal expansion coefficient $\alpha(B)$ are affected by the
fluctuations or localization in a correlated fashion. As we will
see below, the MR measurements in the crossover region can present
indicative data on the quasiparticles availability. Such MR
measurements were carried out in $\rm{ CeCoIn_5}$ when the system
transits from LFL to NFL regime at elevated temperatures and fixed
magnetic fields \cite{pag,mal}.

In this Letter, we analyze MR of $\rm{ CeCoIn_5}$ and show that the
crossover from negative to positive MR at elevated temperatures and
fixed magnetic fields can be well captured utilizing fermion
condensation quantum phase transition (FCQPT) concept based on the
quasiparticles paradigm \cite{khs,obz,ams,volovik}. We demonstrate
that the crossover is regulated by the universal behavior of the
effective mass $M^*(B,T)$ observed in many HF metals. It is
exhibited by $M^*(B,T)$ when HF metal transits from LFL regime
(induced by the application of magnetic field) to NFL one taking
place at rising temperatures. The above behavior of the effective
mass also generates kinks (crossover points from fast to slow
growth at elevated temperatures) in thermodynamic characteristics
(like specific heat, magnetization etc). We show that the above
kink-like peculiarity separates two distinct energy scales - low
temperature LFL scale and high temperature one related to NFL
regime. Our calculations of MR are in good agreement with
observations and allow us to reveal new scaling behavior of both MR
and the kinks.

\section{Scaling behavior of the kinks}

To study universal low temperature features of HF metals, we use the
model of homogeneous heavy-fermion liquid with the effective mass
$M^*(T,B,x)$, where $x=p_F^3/3\pi^2$ is a number density and $p_F$
is Fermi momentum \cite{land}. This model permits to avoid
complications associated with the crystalline anisotropy of solids
\cite{shag1}. We first outline the case when at $T\to 0$ the
heavy-electron liquid behaves as LFL and is brought to the LFL side
of FCQPT by tuning of a control parameter like $x$. At elevated
temperatures the system transits to the NFL state. The dependence
$M^*(T,x)$ is governed by Landau equation \cite{land}
\begin{equation}
\frac{1}{M^*(T,x)}=\frac{1}{M}+\int\frac{{\bf p}_F{\bf p}}{p_F^3}F
({\bf p_F},{\bf p})\frac{\partial n({\bf
p},T,x)}{\partial{p}}\frac{d{\bf p}}{(2\pi)^3},\label{LQ}
\end{equation}
where $n({\bf p},T,x)$ is the distribution function of
quasiparticles and $F({\bf p}_F,{\bf p})$ is Landau interaction
amplitude, $M$ is a free electron mass. At $T=0$, eq. \eqref{LQ}
reads \cite{land} $M^*/M=1/(1-N_0F^1(p_F,p_F)/3)$. Here $N_0$ is
the density of states of a free electron gas,  $F^1(p_F,p_F)$ is
the $p$-wave component of Landau interaction amplitude $F$. Taking
into account that $x =p_F^3/3\pi^2$, we rewrite the amplitude as
$F^1(p_F,p_F)=F^1(x)$. When at some critical point $x=x_{FC}$,
$F^1(x)$ achieves certain threshold value, the denominator tends to
zero and the system undergoes FCQPT related to divergency of the
effective mass \cite{khs,volovik,obz,khodb},
\begin{equation}
\frac{M^*(x)}{M}=A+\frac{B}{x_{FC}-x}\label{AB}.
\end{equation}
Equation \eqref{AB} is valid in both 3D and 2D cases, while the
values of factors $A$ and $B$ depend on the dimensionality. The
approximate solution of Eq. \eqref{LQ} is of the form \cite{shag2}
\begin{eqnarray}
\frac{M}{{M^* (T)}}&=&\frac{M}{{M^*(x)}}+\beta f(0)\ln\left\{
{1+\exp(-1/\beta)}\right\}\nonumber \\
&+&\lambda _1\beta^2+\lambda_2 \beta^4 + ...,\label{zui1}
\end{eqnarray}
where $\lambda_1>0$ and $\lambda_2<0$ are constants of order unity,
$\beta=TM^*(T)/p_F^2$ and $f(0)\sim F^1(x_{FC})$. It follows from
Eq. (\ref{zui1}) that the effective mass $M^*$ as a function of $T$
and $x$ reveals three different regimes at growing temperature. At
the lowest temperatures we have LFL regime with $M^*(T,x)\simeq
M^*(x)+aT^2$ with $a<0$ since $\lambda_1>0$. The effective mass as
a function of $T$ decays down to a minimum and after grows,
reaching its maximum $M^*_M$ at some temperature $T_M(x)$ then
subsequently diminishing as $T^{-2/3}$ \cite{ckz,obz}. Moreover,
the closer is the number density $x$ to its threshold value
$x_{FC}$, the higher is the rate of the growth. The peak value
$M^*_M$ grows also, but the maximum temperature $T_M$ lowers. Near
this temperature the last "traces" of LFL regime disappear,
manifesting themselves in the divergence of above low-temperature
series and substantial growth of $M^*(x)$. Numerical calculations
based on Eqs. (\ref{LQ}) and (\ref{zui1}) show that at rising
temperatures $T>T_{1/2}$ ($T_{1/2}$ is a characteristic temperature
determining the validity of regime \eqref{r2}, see Ref.
\cite{shag2} for details) the linear term $\propto \beta$ gives the
main contribution and leads to new regime when Eq. (\ref{zui1})
reads $M/M^*(T)\propto\beta $ yielding
\begin{equation}\label{r2}
    M^*(T) \propto T^{-1/2}.
\end{equation} We remark that Eq. \eqref{r2} ensures that at $T\geq
T_{1/2}$ the resistivity behaves as $\rho(T)\propto T$ \cite{obz}.
\begin{figure}[!ht]
\begin{center}
\vspace*{-0.5cm}
\includegraphics [width=0.48\textwidth]{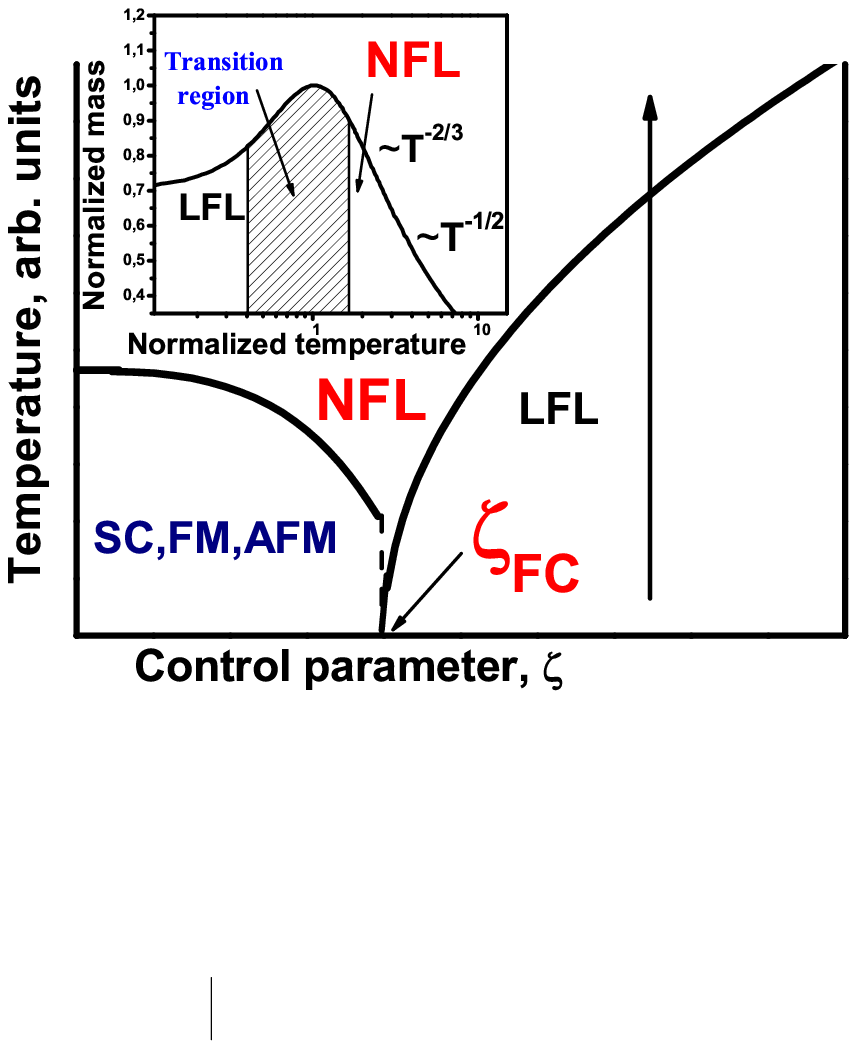}
\vspace*{-4.0cm}
\end{center}
\caption{Schematic phase diagram of the systems under consideration.
Control parameter $\zeta$ represents number density (or doping) $x$,
magnetic field $B$, pressure $P$ etc. $\zeta_{\rm FC}$ denotes the
point of effective mass divergence. $\rm {SC,FM,AFM}$ denote the
superconducting, ferromagnetic and antiferromagnetic states,
respectively. The vertical arrow shows LFL-NFL transitions at rising
temperatures and fixed $\zeta$. Inset shows a schematic plot of the
normalized effective mass $M^*_N=M^*(T/T_M)/M^*_M$ ($M^*_M$ is its
maximal value at $T=T_M$) versus the normalized temperature
$T_N=T/T_M$. Several regions are shown. First goes the LFL regime
($M^*_N(T_N)\sim$ const) at $T_N\ll 1$, then the transition regime
(the hatched area) where $M^*_N$ reaches its maximum. At elevated
temperatures $T^{-2/3}$ regime occurs followed by $T^{-1/2}$
behavior, see Eq. (\ref{r2}).}\label{MT}
\end{figure}
Near the critical point $x_{FC}$ ($M/M^*(x\to x_{FC})\to 0$), the
behavior of the effective mass changes dramatically since the first
term in the right-hand side of Eq. \eqref{zui1} vanishes so that
the second term becomes dominant. As a result, we can no more
measure the mass $M^*$ in units of $M$ (as $M/M^*(x\to x_{FC})\to
0$) and we have to measure $M^*$ in units of $M^*_M$ and $T$ in
units of $T_M$. Latter scales can be viewed as natural ones.

The schematic phase diagram of HF liquid is reported Fig. \ref{MT}.
The control parameter $\zeta $ can be pressure $P$, magnetic field
$B$, or doping (density) $x$. At $\zeta=\zeta_{FC}$, FCQPT takes
place leading to a strongly degenerated state. This state is
captured by the superconducting (SC), ferromagnetic (FM),
antiferromagnetic (AFM) etc. states lifting the degeneracy
\cite{obz}. The variation of $\zeta$ drives the system from NFL
region to LFL one. For example, in the case of magnetic field $B$,
$\zeta_{FC}=B_{c0}$, where $B_{c0}$ is a critical magnetic field,
such that at $B>B_{c0}$ the system is driven towards its LFL regime.
Below we consider the case with $\zeta>\zeta_{FC}$ when the system
is on the LFL side of FCQPT. The inset demonstrates the behavior of
the normalized effective mass $M^*_N=M^*/M^*_M$ versus normalized
temperature $T_N=T/T_M$. Both $T^{-2/3}$ and $T^{-1/2}$ regimes are
marked as NFL ones since the effective mass depends strongly on
temperature. The temperature region $T\simeq T_M$ signifies the
crossover between the LFL regime with almost constant effective mass
and NFL behavior, given by $T^{-2/3}$ dependence. Thus temperatures
$T\sim T_M$ can be regarded as the crossover region between LFL and
NFL regimes.

It turns out that $M^*(T,x)$ in the entire $T\leq T_{1/2}$ range can
be well approximated by a simple universal interpolating function
\cite{obz,shag2,ckz}. The interpolation occurs between the LFL
($M^*\propto T^2$) and NFL ($M^*\propto T^{-1/2}$, see Eq.
\eqref{r2}) regimes thus describing the above crossover. Introducing
the dimensionless variable $y=T_N=T/T_M$, we obtain the desired
expression
\begin{equation}
\frac{M^*(T/T_M,x)}{M^*_M} = {M^*_N(y)}\approx
\frac{M^*(x)}{M^*_M}\frac{1+c_1y^2}{1+c_2y^{8/3}}. \label{UN2}
\end{equation}
Here $M^*_N(y)$ is the normalized effective mass,  $c_1$ and $c_2$
are parameters, obtained from the condition of best fit to
experiment. To correct the behavior of $M^*_N(y)$ at rising
temperatures $M^*\sim T^{-1/2}$, we add a term to Eq. \eqref{UN2}
and obtain
\begin{equation}
M^*_N(y)\approx\frac{M^*(x)}{M^*_M}\left[\frac{1+c_1y^2}{1+c_2y^{8/3}}
+c_3\frac{\exp(-1/y)}{\sqrt{y}}\right], \label{HC28}\end{equation}
where $c_3$ is a parameter. The last term on the right hand side of
Eq. \eqref{HC28} makes $M^*_N$ satisfy Eq. \eqref{r2} at
temperatures $T/T_M>2$.

At small magnetic fields $B$ (that means that Zeeman splitting is
small), the effective mass does not depend on spin variable and $B$
enters Eq. \eqref{LQ} as $B\mu_B/T$ ($\mu_B$ is Bohr magneton)
making $T_M\propto B\mu_B$ \cite{ckz,obz,shag2}. The application of
magnetic field restores the LFL behavior, and at $T\leq T_M$ the
effective mass depends on $B$ as \cite{ckz,obz}
\begin{equation}\label{B32}
M^*(B)\propto (B-B_{c0})^{-2/3}.
\end{equation}
Note that in some cases $B_{c0}=0$. For example, the HF metal $\rm{
CeRu_2Si_2}$is characterized by $B_{c0}=0$ and shows neither
evidence of the magnetic ordering or superconductivity nor the LFL
behavior down to the lowest temperatures \cite{takah}. In our
simple model $B_{c0}$ is taken as a parameter. We conclude that
under the application of magnetic field the variable
\begin{equation}\label{YTB}
y=T/T_M\propto \frac{T}{\mu_B(B-B_{c0})}
\end{equation}
remains the same and the normalized effective mass is again
governed by Eqs. \eqref{UN2} and \eqref{HC28} which are the final
result of our analytical calculations. We note that the obtained
results are in agreement with numerical calculations
\cite{obz,ckz}.

The normalized effective mass $M^*_N(y)$ can be extracted from
experiments on HF metals. For example, $M^*(T,B)\propto
C(T)/T\propto S(T)/T\propto \chi_{AC}(T)$, where $S(T)$ is the
entropy, $C(T)$ is the specific heat and $\chi_{AC}(T)$ is ac
magnetic susceptibility \cite{ckz,obz}. If the corresponding
measurements are carried out at fixed magnetic field $B$ (or at
fixed $x$ and $B$) then, as it seen from Fig. \ref{MT}, the
effective mass reaches the maximum at some temperature $T_M$. Upon
normalizing both the effective mass by its peak value $M^*_M$ at
each field $B$  and the temperature by $T_M$, we observe that all
the curves merge into a single one, given by Eqs. (\ref{UN2}) and
\eqref{HC28} thus demonstrating a scaling behavior.

\begin{figure} [! ht]
\begin{center}
\vspace*{-0.5cm}
\includegraphics [width=0.47\textwidth]{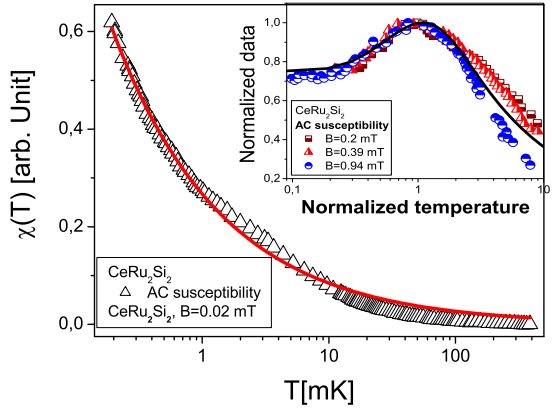}
\vspace*{-1.0cm}
\end{center}
\caption{Temperature dependence of the $ac$ susceptibility
$\chi_{AC}$  for $\rm{ CeRu_2Si_2}$. The solid curve is a fit for
the data shown by the triangles at $B=0.02$ mT and represented by
the function $\chi(T)=a/\sqrt{T}$ given by Eq. \eqref{r2} with $a$
being a fitting parameter. Inset shows the normalized effective
mass versus normalized temperature $y$ extracted from $\chi_{AC}$
measured at different fields as indicated in the inset
\cite{takah}. The solid curve traces the universal behavior of
$M^*_N(y)$ determined by Eq. (\ref{UN2}). Parameters $c_1$ and
$c_2$ are adjusted to fit the average behavior of the normalized
effective mass $M^*_N(y)$.}\label{MRM}
\end{figure}

To verify Eq. \eqref{r2}, we use measurements of $\chi_{AC}(T)$ in
$\rm{ CeRu_2Si_2}$ at $B=0.02$ mT at which this HF metal
demonstrates the NFL behavior \cite{takah}. It is seen from Fig.
\ref{MRM} that Eq. \eqref{r2} gives good description of the facts
in the extremely wide range of temperatures. The inset to Fig.
\ref{MRM} exhibits a fit for $M^*_N(y)$ extracted from measurements
of $\chi_{AC}(T)$ at different magnetic fields, clearly indicating
that the function given by Eq. \eqref{UN2} represents a good
approximation for $M^*_N(y)$ when the system transits from the LFL
regime to NFL one.

\begin{figure} [! ht]
\begin{center}
\vspace*{-0.8cm}
\includegraphics [width=0.49\textwidth]{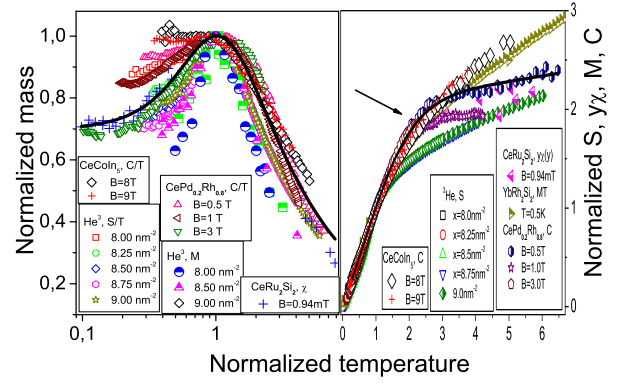}
\end{center}
\vspace*{-0.8cm} \caption{The left panel. The normalized effective
mass $M^*_N$ versus the normalized temperature $y=T/T_M$. The
dependence $M^*_N (y)$ is extracted from measurements of $S(T)/T$
and magnetization $M$ on 2D $\rm{ ^3He}$ \cite{he3}), from $ac$
susceptibility $\chi_{AC}(T)$ collected on $\rm{ CeRu_2Si_2}$
\cite{takah} and from $C(T)/T$ collected on $\rm{ CePd_{1-x}Rh_x}$
\cite{pikul}. The data are collected for different densities and
magnetic fields shown in the left bottom corner. The solid curve
traces the universal behavior of the normalized effective mass
determined by Eq. (\ref{UN2}). Parameters $c_1$ and $c_2$ are
adjusted for $\chi_{N}(T_N,B)$ at $B=0.94$ mT. The right panel. The
normalized specific heat $C(y)$ of $\rm{ CePd_{1-x}Rh_x}$ at
different magnetic fields $B$, normalized entropy $S(y)$ of $\rm{
^3He}$ at different number densities $x$, and the normalized
$y\chi(y)$ at $B=0.94$ mT versus normalized temperature $y$ are
shown. The upright triangles depict the normalized `average'
magnetization $M+B\chi$ collected on $\rm{ YbRu_2Si_2}$ \cite{steg}.
The kink (shown by the arrow) in all the data is clearly seen in the
transition region $y\geq 1$. The solid curve represents $yM^*_N(y)$
with parameters $c_1$ and $c_2$ adjusted for magnetic susceptibility
of $\rm{ CeRu_2Si_2}$ at $B=0.94$ mT.}\label{f2}
\end{figure}

$M^*_N(y)$ extracted from the entropy $S(T)/T$ and magnetization
$M$ measurements on the $^3$He film \cite{he3} at different
densities $x$ is reported in the left panel of Fig. \ref{f2}. In
the same panel, the data extracted from the heat capacity of the
ferromagnet $\rm{ CePd_{0.2}Rh_{0.8}}$ \cite{pikul} and the AC
magnetic susceptibility of the paramagnet $\rm{ CeRu_2Si_2}$
\cite{takah} are plotted for different magnetic fields. It is seen
that the universal behavior of the normalized effective mass given
by Eq. (\ref{UN2}) and shown by the solid curve is in accord with
the experimental facts. All 2D $\rm{ ^3He}$ substances are located
at $\zeta>\zeta_{FC}$ (see Fig. \ref{MT}), where the system
progressively disrupts its LFL behavior at elevated temperatures.
In that case the control parameter, driving the system towards its
quantum critical point (QCP) is merely the number density $x$. It
is seen that the behavior of $M^*_N(y)$, extracted from $S(T)/T$
and magnetization $M$ of 2D $\rm{ ^3He}$ looks very much like that
of 3D HF compounds. In the right panel of Fig. \ref{f2}, the
normalized data on $C(y)$, $S(y)$, $y\chi(y)$ and $M(y)+y\chi(y)$
extracted from data collected on $\rm{ CePd_{1-x}Rh_x}$
\cite{pikul} , $\rm{ ^3He}$ \cite{he3}, $\rm{ CeRu_2Si_2}$
\cite{takah}, $\rm CeCoIn_5$ \cite{bian} and $\rm{ YbRu_2Si_2}$
\cite{steg} respectively are presented. Note that in the case of
$\rm{ YbRu_2Si_2}$, the variable $y=(B-B_{c0})\mu_B/T_M$ can be
viewed as effective normalized temperature. As seen from Eq.
\eqref{UN2}, this representation of the variable $y$ is correct
when the temperature is a fixed parameter.

It is seen from the right panel of Fig. \ref{f2} that all the data
exhibit the kink (shown by arrow) at $y\geq 1$ taking place as soon
as the system enters the transition region from the LFL regime to
the NFL one and corresponding to the temperatures where the
vertical arrow in Fig. \ref{MT} crosses the solid line. It is also
seen that the low temperature LFL scale of the thermodynamic
functions (as a function of $y$) is characterized by the fast
growth and the high temperature one related to the NFL behavior is
characterized by the slow growth. As a result, we can identify the
energy scales near QCP, discovered in Ref. \cite{steg}: the
thermodynamic characteristics exhibit the kinks (crossover points
from the fast to slow growth at elevated temperatures) which
separate the low temperature LFL scale and high temperature one
related to NFL regime.

\section{Scaling behavior of the magnetoresistance}
By definition, MR is given by
\begin{equation}
\rho_{mr}(B,T)=\frac{\rho(B,T)-\rho(0,T)}{\rho(0,T)},\label{HC23}
\end{equation}
We apply Eq. (\ref{HC23}) to study MR of strongly correlated
electron liquid versus temperature $T$ as a function of magnetic
field $B$. The resistivity $\rho(B,T)$ is
\begin{equation}
\rho(B,T)=\rho_0+\Delta\rho(B,T)+\Delta\rho_{L}(B,T),
\label{RBT}
\end{equation}
where $\rho_0$ is a residual resistance, $\Delta\rho=c_1AT^2$, $c_1$
is a constant, $A$ is a coefficient determining the temperature
dependence of the resistivity $\rho=\rho_0+AT^2$. The classical
contribution $\Delta\rho_{L}(B,T)$ to MR due to orbital motion of
carriers induced by the Lorentz force obeys the Kohler's rule
\cite{zim}. We note that $\Delta\rho_{L}(B)$ $\ll\rho(0, T)$ as it
is assumed in the weak-field approximation. To calculate $A$, we use
the quantities $\gamma_0=C/T\propto M^*$ and/or $\chi\propto M^*$ as
well as employ the fact that Kadowaki-Woods ratio
$K=A/\gamma_0^2\propto A/\chi^2=const$ \cite{kadw}. As a result, we
obtain $A\propto (M^*)^2$ \cite{kadw,kh_z,tky}, so that
$\Delta\rho(B,T)=c(M^*(B,T))^2T^2$ and $c$ is a constant. Suppose
that the temperature is not very low, so that $\rho_0\leq
\Delta\rho(B=0,T)$, and $B\geq B_{c0}$. Substituting (\ref{RBT})
into (\ref{HC23}), we find that \cite{shag_mr}
\begin{equation}
\rho_{mr}\simeq \frac{\Delta\rho_{L}}{\rho}+cT^2
\frac{(M^*(B,T))^2-(M^*(0, T))^2}{\rho(0, T)}\label{HC25}.
\end{equation}

Consider the qualitative behavior of MR described by Eq.
(\ref{HC25}) as a function of $B$ at a certain temperature $T=T_0$.
In weak magnetic fields, when $T_0\geq T_{1/2}$ and the system
exhibits NFL regime (see Fig. \ref{MT}), the main contribution to
MR is made by the term $\Delta\rho_{L}(B)$, because the effective
mass is independent of the applied magnetic field. Hence, $|M^*(B,
T)-M^*(0,T)|/M^*(0, T)\ll1$ and the leading contribution is made by
$\Delta\rho_{L}(B)$. As a result, MR is an increasing function of
$B$. When $B$ becomes so high that $T_M(B)\sim \mu_B(B-B_{c0})\sim
T_0$, the difference $(M^*(B, T)-M^*(0, T))$ becomes negative and
MR as a function of $B$ reaches its maximal value at $T_M(B)\sim
T_0$ when the kink occurs, see the right panel of Fig. \ref{f2}. At
further increase of magnetic field, when $T_M(B)>T_0$, the
effective mass $M^*(B,T)$ becomes a decreasing function of $B$, as
follows from Eq. (\ref{B32}). As $B$ increases,
\begin{equation}
\frac{(M^*(B,T)-M^*(0,T))}
{M ^*(0, T)}\to -1,\label{HC25a}
\end{equation}
and the magnetoresistance, being a decreasing function of $B$, is
negative.

Now we study the behavior of MR as a function of $T$ at fixed value
$B_0$ of magnetic field. At low temperatures $T\ll T_M(B_0)$, it
follows from Eqs. (\ref{UN2}) and (\ref{B32}) that
$M^*(B_0,T)/M^*(0,T)\ll1$, and it is seen from Eq. \eqref{HC25a}
that $\rho_{mr}(B_0,T)\sim-1$, because
$\Delta\rho_{L}(B_0,T)/\rho(0,T)\ll1$. We note that $B_0$ must be
relatively high to guarantee that $M^*(B_0,T)/M^*(0,T)$ $\ll1$. As
the temperature increases, MR increases, remaining negative. At
$T\simeq T_M(B_0)$, MR is approximately zero, because
$\rho(B_0,T)\simeq\rho(0,T)$ at this point. This allows us to
conclude that the change of the temperature dependence of
resistivity $\rho(B_0,T)$ from quadratic to linear manifests itself
in the transition from negative to positive MR. One can also say
that the transition takes place when the kink occurs (as shown by
the arrow in the right panel of Fig. \ref{f2}) and the system goes
from the LFL behavior to the NFL one. At $T\geq T_M(B_0)$, the
leading contribution to MR is made by $\Delta\rho_{L}(B_0,T)$ and MR
reaches its maximum. At $T_M(B_0)\ll T$, MR is a decreasing function
of the temperature, because
\begin{equation}
\frac{|M^*(B,T)-M^*(0, T)|}{M^*(0,T)}
\ll1,\label{HC25b}
\end{equation}
and $\rho_{mr}(B_0,T)\ll1$. Both transitions (from positive to
negative MR with increasing $B$ at  fixed temperature $T$ and from
negative to positive MR with increasing $T$ at fixed $B$ value) have
been detected in measurements of the resistivity of $\rm{ CeCoIn_5}$
in a magnetic field \cite{pag}.
\begin{figure} [! ht]
\begin{center}
\vspace*{-0.5cm}
\includegraphics [width=0.47\textwidth]{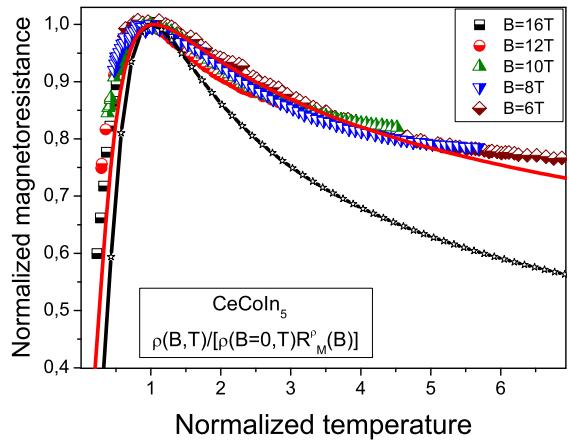}
\end {center}\vspace{-0.8cm}
\caption {The normalized magnetoresistance $R^{\rho}_N(y)$ given by
Eq. \eqref{HC27} versus normalized temperature $y=T/T_{\rm Rm}$.
$R^{\rho}_N(y)$ was extracted from MR shown in Fig. \ref{MRT} and
collected on $\rm{ CeCoIn_5}$ at fixed magnetic fields $B$
\cite{pag} listed in the right upper corner. The starred  line
represents our calculations based on Eqs. \eqref{UN2} and
\eqref{HC27} with the parameters extracted from $ac$ susceptibility
of $\rm CeRu_2Si_2$ (see the caption to Fig. \ref{MRM}). The solid
line displays our calculations based on Eqs. \eqref{HC28} and
\eqref{HC27}; only one parameter was used to fit the data, while
the other were extracted from the $ac$ susceptibility measured on
$\rm{ CeRu_2Si_2}$.}\label{MRTU}
\end{figure}

Let us turn to quantitative analysis of MR. As it was mentioned
above, we can safely assume that the classical contribution
$\Delta\rho_{L}(B,T)$ to MR is small as compared to
$\Delta\rho(B,T)$. Omission of $\Delta\rho_{L}(B,T)$ allows us to
make our analysis and results transparent and simple since the
behavior of $\Delta\rho_{L}(B_0,T)$ is not known in the case of HF
metals. Consider the ratio $R^{\rho}=\rho(B,T)/\rho(0,T)$ and assume
for a while that the residual resistance $\rho_0$ is small in
comparison with the temperature dependent terms. Taking into account
Eq. \eqref{RBT} and $\rho(0,T)\propto T$, we obtain from Eq.
\eqref{HC25}
\begin{equation}
R^{\rho}=\rho_{mr}+1=
\frac{\rho(B,T)}{\rho(0, T)}\propto T(M^*(B,T))^2.\label{HC26}
\end{equation}
It follows from Eqs.  \eqref{UN2} and \eqref{HC26} that the ratio
$R^{\rho}$ reaches its maximal value $R^{\rho}_M$ at some
temperature $T_{\rm Rm}\sim T_M$. If the ratio is measured in units
of its maximal value $R^{\rho}_M$ and $T$ is measured in units of
$T_{\rm Rm}\sim T_M$ then it is seen from Eqs. \eqref{UN2},
\eqref{HC28} and \eqref{HC26} that the normalized MR
\begin{equation}
R^{\rho}_N(y)=
\frac{R^{\rho}(B,T)}{R^{\rho}_M(B)}\simeq y(M^*_N(y))^2\label{HC27}
\end{equation}
becomes a universal function of the only variable $y=T/T_{\rm Rm}$.
To verify Eq. \eqref{HC27}, we use MR obtained in measurements on
CeCoIn$_5$, see Fig. 1(b) of Ref. \cite{pag}. The results of the
normalization procedure of MR are reported in Fig. \ref{MRTU}. It is
clearly seen that the data collapse into the same curve, indicating
that the normalized magnetoresistance $R^{\rho}_N$ well obeys the
scaling behavior given by Eq. \eqref{HC27}. This scaling behavior
obtained directly from the experimental facts is a vivid evidence
that MR behavior is predominantly governed by the effective mass
$M^*(B,T)$.

Now we are in position to calculate $R^{\rho}_N(y)$ given by Eq.
\eqref{HC27}. Using Eq. \eqref{UN2} to parameterize $M^*_N(y)$, we
extract parameters $c_1$ and $c_2$ from measurements of the magnetic
$ac$ susceptibility $\chi$ on $\rm CeRu_2Si_2$ \cite{takah} and
apply Eq. \eqref{HC27} to calculate the normalized ratio. It is seen
that the calculations shown by the starred line in Fig. \ref{MRTU}
start to deviate from experimental points at elevated temperatures.
To improve the coincidence, we employ Eq. \eqref{HC28} which
describes the behavior of the effective mass at elevated
temperatures in accord with Eq. \eqref{r2} and ensures that at these
temperatures the resistance behaves as $\rho(T)\propto T$. In Fig.
\ref{MRTU}, the fit of $R^{\rho}_N(y)$ by Eq. \eqref{HC28} is shown
by the solid line. Constant $c_3$ is taken as a fitting parameter,
while the other were extracted from $ac$ susceptibility of $\rm{
CeRu_2Si_2}$ as described in the caption to Fig. \ref{MRM}.
\begin{figure} [! ht]
\begin{center}
\vspace*{-0.5cm}
\includegraphics [width=0.40\textwidth]{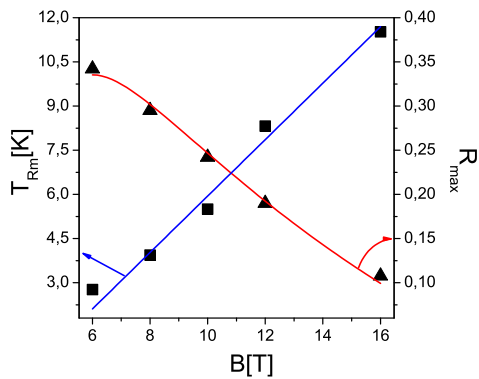}
\end {center}\vspace{-1.0cm}
\caption {The peak temperatures $T_{\rm Rm}$ (squares) and the peak
values $R_{\rm max}$ (triangles) versus magnetic field $B$
extracted from measurements of MR \cite{pag}. The solid lines
represent our calculations based on Eqs. \eqref{RTB1} and
\eqref{RTB2}.}\label{TB}
\end{figure}

Before discussing  the magnetoresistance $\rho_{mr}(B,T)$ given by
Eq. \eqref{HC23}, we consider the magnetic field dependencies of
both the MR peak value $R_{\rm max}(B)$ and corresponding peak
temperature $T_{\rm Rm}(B)$. It is possible to use Eq. \eqref{HC26}
which relates the position and value of the peak with the function
$M^*(B,T)$. Since $T_{\rm Rm}\propto \mu_BB$,  $B$ enters Eq.
\eqref{HC26} only as tuning parameter of QCP, as both $\Delta\rho_L$
and $\rho_0$ were omitted. At $B\to B_{c0}$ and $T\leq T_{\rm
Rm}(B)$, this omission is not correct since $\Delta\rho_L$ and
$\rho_0$ become comparable with $\Delta\rho(B,T)$. Therefore, both
$R_{\rm max}(B)$ and $T_{\rm Rm}(B)$ are not characterized by any
critical field, being a continuous function at the quantum critical
filed $B_{c0}$, in contrast to $M^*(B,T)$ which peak value diverges
and the peak temperature tends to zero at $B_{c0}$ as it follows
from Eqs. \eqref{B32} and \eqref{YTB}. Thus, we have to take into
account $\Delta\rho_{L}(B,T)$ and $\rho_0$ which prevent $T_{\rm
Rm}(B)$ from vanishing and make $R_{\rm max}(B)$ finite at $B\to
B_{c0}$. As a result, we have to replace $B_{c0}$ by some effective
field $B_{eff}<B_{c0}$ and take $B_{eff}$ as a parameter which
imitates the contributions coming from both $\Delta\rho_{L}(B,T)$
and $\rho_0$. Upon modifying Eq. \eqref{HC26} by taking into account
$\Delta\rho_{L}(B,T)$ and $\rho_0$, we obtain
\begin{eqnarray}
&&T_{\rm Rm}(B)\simeq b_1(B-B_{eff}),\label{RTB1}\\
&&R_{\rm max}(B)\simeq
\frac{b_2(B-B_{eff})^{-1/3}-1}{b_3(B-B_{eff})^{-1}+1}.\label{RTB2}
\end{eqnarray}
Here $b_1$, $b_2$, $b_3$ and $B_{eff}$ are fitting parameters. It is
pertinent to note that while deriving Eq. \eqref{RTB2} we use Eq.
\eqref{RTB1} with substitution $(B-B_{eff})$ for $T$. Then, Eqs.
\eqref{RTB1} and \eqref{RTB2} are not valid at $B\lesssim B_{c0}$.
In Fig. \ref{TB}, we show the field dependence of both $T_{\rm Rm}$
and $R_{\rm max}$, extracted from measurements of MR \cite{pag}. It
is seen that both $T_{\rm Rm}$ and $R_{\rm max}$ are well described
by Eqs. \eqref{RTB1} and \eqref{RTB2} with $B_{eff}=$3.8 T. We note
that this value of $B_{eff}$ is in good agreement with observations
obtained from the $B-T$ phase diagram of $\rm{ CeCoIn_5}$, see the
position of the MR maximum shown by the filled circles in Fig. 3 of
Ref. \cite{pag}.
\begin{figure} [! ht]
\begin{center}
\vspace*{-0.6cm}
\includegraphics [width=0.47\textwidth]{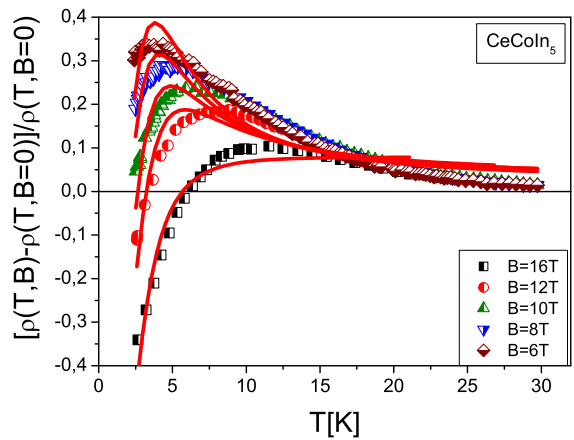}
\end {center}\vspace{-1.0cm}
\caption {MR versus temperature $T$ as a function of magnetic field
$B$. The experimental data on MR were collected on $\rm{ CeCoIn_5}$
at fixed magnetic field $B$ \cite{pag} shown in the right bottom
corner of the Figure. The solid lines represent our calculations,
Eq. \eqref{UN2} is used to fit the effective mass entering Eq.
\eqref{HC27}.}\label{MRT}
\end{figure}

To calculate $\rho_{mr}(B,T)$, we apply Eq. \eqref{HC27} to describe
its universal behavior, Eq. \eqref{UN2} for the effective mass along
with Eqs. \eqref{RTB1} and \eqref{RTB2} for MR parameters. Figure
\ref{MRT} shows the calculated MR versus temperature as a function
of magnetic field $B$ together with the experimental points from
Ref. \cite{pag}. We recall that the contributions coming from
$\Delta\rho_{L}(B,T)$ and $\rho_0$ were omitted. As seen from Fig.
\ref{MRT}, our description of experiment is pretty good.

\section{Summary}

Our comprehensive theoretical study of MR shows that it is (similar
to other thermodynamic characteristics like magnetic
susceptibility, specific heat etc) governed by the scaling behavior
of the quasiparticle effective mass. The crossover from negative to
positive MR occurs at elevated temperatures and fixed magnetic
fields when the system transits from the LFL behavior to NFL one
and can be well captured by this scaling behavior. This behavior
permits to identify the energy scales near QCP, discovered in Ref.
\cite{steg}. Namely, the thermodynamic characteristics (like
specific heat, magnetization etc) consist of the low temperature
LFL scale characterized by the fast growth and the high temperature
one related to the NFL behavior and characterized by the slow
growth. These scales are separated by the kinks in the transition
region. Obtained theoretical results are in good agreement with
experimental facts and allow us to reveal for the first time a new
scaling behavior of both magnetoresistance and kinks separating the
different energy scales.

\section{Acknowledgements}

This work was supported in part by the grants: RFBR No.
09-02-00056, DOE and NSF No. DMR-0705328, and the Hebrew University
Intramural Funds.

\end{document}